\documentclass{jpp}
\usepackage{graphicx}

\usepackage[utf8]{inputenc}
\usepackage[T1]{fontenc}
\usepackage{amsmath}
\usepackage{siunitx}
\usepackage{comment}
\usepackage{url}

\newcommand{\deriv}[2]{\frac{\mathrm{d}#1}{\mathrm{d}#2}}
\newcommand{\rmc}{\mathrm{c}}
\newcommand{\rmd}{\mathrm{d}}
\newcommand{\rme}{\mathrm{e}}
\newcommand{\EC}{E_\mathrm{c}}

\newcommand{\nubar}[1]{\bar{\nu}_\mathrm{#1}}
\newcommand{\Zeff}{Z_\mathrm{eff}}
\newcommand{\EceffNorm}{\bar{E}_\rmc^\mathrm{eff}}
\newcommand{\lnLc}{\ln{\rmLambda_\rmc}}

\DeclareMathOperator{\arctanh}{arctanh}
\DeclareMathOperator{\Heavi}{H}

\shorttitle{Criterion for runaway electron generation in activated tokamaks}
\shortauthor{B.~Zaar,  I.~Pusztai, I.~Ekmark, and T.~Fülöp}

\title{An analytical criterion for significant runaway electron generation in activated tokamaks}

\author{B.~Zaar\aff{1}
  \corresp{\email{bjorn.zaar@chalmers.se}},  
     I.~Pusztai\aff{1}, I.~Ekmark\aff{1}, \and T.~Fülöp\aff{1}
}
\affiliation{
\aff{1}Department of Physics and Astronomy, Chalmers University of Technology, Gothenburg, SE-41296, Sweden
}

\begin{document}

\maketitle

\begin{abstract}
A disrupting plasma in a high-performance tokamak such as ITER or SPARC may generate large runaway electron currents that, upon impact with the tokamak wall, can cause serious damage to the device. To quickly identify regions of safe operation in parameter space, it is useful to develop reduced models and analytical criteria that predict when a significant fraction of the Ohmic current is converted into a current of runaway electrons. In deuterium-tritium plasmas, the seed runaway current may have a significant contribution from --- or may even be dominated by --- tritium beta decay and Compton scattering. In this work, a criterion for significant runaway electron generation that includes tritium beta decay and Compton scattering sources is developed. The avalanche gain factor includes the effects of partial screening of injected noble gases. The result is an analytical model that can predict significant runaway electron generation in the next generation of activated tokamak devices. The model is validated by fluid simulations using \textsc{Dream} \citep{DREAM} and is shown to delineate regions in parameter space where significant runaway electron generation may occur. 
\end{abstract}

\section{Introduction}

Plasma-terminating disruptions are one of the crucial problems facing magnetic fusion devices with large plasma currents, such as ITER \citep{Hender2007,Bandyopadhyay2025}. Such events lead to a sudden cooling of the plasma --- a thermal quench. The thermal quench is associated with a substantial increase in the plasma resistivity, causing the current to decay, which induces a strong toroidal electric field. If the electric field exceeds a critical value, above which the friction force from collisions and radiation becomes smaller than the accelerating force from the electric field for electrons in the tail of the bulk Maxwellian distribution, these electrons may run away. In such cases, a large part of the initial plasma current can be converted into a beam of relativistic runaway electrons \citep{Breizman2019}.

Runaway electrons can be produced by a range of mechanisms. First, electrons can enter the runaway region in momentum space through the diffusive leak from the thermal population at a rate that is exponentially sensitive to the electric field (so-called Dreicer generation) \citep{Connor1975}. Second, in the case of sudden cooling, when the collision frequency is lower than the cooling rate, fast electrons do not have time to thermalize, and a hot tail forms in the electron distribution \citep{Helander04electron}. If the hot tail electrons have energies above the critical energy, they may run away. Finally, during deuterium-tritium (DT) operation, fast electrons can be produced by nuclear reactions. Tritium ions undergoing beta decay produce electrons in the keV range, and Compton scattering of gamma rays from the activated wall can knock both free and bound electrons into the runaway regime \citep{MartinSolis2017}. Collectively, these four mechanisms are referred to as primary generation mechanisms.

Furthermore, and perhaps most importantly for reactor-scale fusion devices, runaway electrons already present in the plasma can create new ones through close collisions with thermal electrons. This leads to an exponential growth in the number of runaway electrons --- an avalanche \citep{Jayakumar1993}. This secondary generation is proportional to the density of existing runaway electrons, and the dynamics become highly nonlinear. Therefore, small variations in the balance between runaway electron generation and transport lead to large differences in the maximum runaway electron current. In addition, the avalanche gain is exponentially sensitive to the initial plasma current and is therefore expected to be significantly larger in future devices with larger plasma currents \citep{Rosenbluth1997}.

Currently envisaged disruption mitigation methods are based on the injection of massive amounts of material \citep{Hollmann2015}, resulting in a cold and partially ionized plasma. In partially ionized plasmas, the nucleus is only partially screened by the bound electron cloud; an energetic electron can penetrate the electron cloud and experience a portion of the normally screened nuclear charge. The effect of this partial screening is expected to be substantial in impurity rich plasmas \citep{Hesslow2019}. In addition, the injection of highly radiating impurities such as neon or argon may lead to even higher avalanche growth rates due to the large number of available target electrons. Therefore, energetic electrons resulting from disruptions can pose a severe risk to future fusion devices.

In order to design operating scenarios for next-generation fusion devices, advanced first-principles simulations are needed. Self-consistent simulations that take into account the evolution of plasma parameters across the entire plasma are computationally expensive, and optimization of operating scenarios based on such simulations is a monumental task. Although simplified fluid models of runaway electron generation will not provide a quantitative prediction of the expected electron energy spectrum, they can be useful in indicating whether large runaway currents are to be expected. 

Based on the approximate solution of two coupled differential equations for the runaway electron density and the inductive electric field, an analytical criterion for significant runaway electron generation was first derived by \citet{Helander2002} and later refined by \citet{Fulop2009} by including the hot-tail runaway seed. Such criteria are useful complements to more comprehensive disruption or runaway electron models, as they can be used to quickly estimate where in parameter space large runaway currents can be expected before committing computational resources to a detailed parameter scan. They can also be used as a "trigger" in integrated modelling code suites to determine whether to execute a more sophisticated model. However, in previous criteria, energetic electrons from nuclear sources (tritium decay and Compton scattering) and the effect of partial screening were not included; hence, they were not applicable for disruptions in activated scenarios mitigated with material injection. 

The focus of this paper is to derive an analytical criterion for significant runaway electron generation in activated scenarios. In \S \ref{sec:fluidEquations}, the zero-dimensional fluid model for runaway electron generation, on which the criterion is based, is presented. \S \ref{sec:avalanche} and \S \ref{sec:nuclearsources} describe the avalanche gain and nuclear runaway electron sources, respectively. The criterion is validated by comparisons with fluid simulations using the \textsc{Dream} code in \S \ref{sec:criterion}, and the conclusions are summarized in \S \ref{sec:conclusions}.

\section{Zero-dimensional fluid model for runaway electron generation} \label{sec:fluidEquations}

The density of the runaway electron population \(n_\mathrm{r}\) is determined by the sum of the primary (seed) and secondary (avalanche) generation mechanisms as 
\begin{equation}
    \deriv{n_\mathrm{r}}{t} = \sum_\mathrm{seeds} \gamma_\mathrm{seed} + \Gamma_\mathrm{ava} n_\mathrm{r}, \label{eq:runawayGeneration}
\end{equation}
with the total seed generation rate (in principle consisting of Dreicer, hot-tail, tritium decay, and Compton scattering sources) \(\sum_\mathrm{seeds}\gamma_\mathrm{seed}\) and the avalanche growth rate \(\Gamma_\mathrm{ava}\). Here, it is assumed that no runaway electrons are present in the plasma at the time of the disruption onset (at \(t = 0\)). All runaway electron generation mechanisms depend on the parallel electric field \(E_\parallel\), which is inductive and is given by
\begin{equation}
	E_{\parallel} = E_{\parallel 0} - \frac{L}{2 \upi R_0} \deriv{I_\mathrm{p}}{t},\label{eq:inductiveElectricField} 
\end{equation}
where
\begin{equation}
    L \approx \mu_0 R_0\left[ \ln{\left( \frac{8R_0}{a} \right)} - 2\right]
\end{equation}
is the self-inductance of the plasma in the large aspect ratio limit, with the vacuum permeability \(\mu_0\), the plasma major radius \(R_0\), and the plasma minor radius \(a\). \(E_{\parallel 0}\) is the contribution to the electric field due to the external transformer. The plasma current \(I_\mathrm{p}\) consists of two contributions, namely the Ohmic current and the runaway electron current, and can be written as
\begin{equation}
	I_\mathrm{p} = \sigma E_\parallel A + n_\mathrm{r} e c A, \label{eq:plasmaCurrent}
\end{equation}
where \(\sigma\) is the Spitzer conductivity, and \(A\) is the (effective) cross-sectional area of the plasma. The constants \(e\) and \(c\) denote the electron charge and the speed of light, respectively. Following the procedure proposed by \citet{Helander2002}, equations \eqref{eq:runawayGeneration}, \eqref{eq:inductiveElectricField}, and \eqref{eq:plasmaCurrent} can be renormalized by introducing the dimensionless quantities
\begin{align}
    n & = \frac{n_\mathrm{r} e c}{j_0}, \\
    E & = \frac{E_\parallel}{\EC}, \\
    t' & = \frac{t}{\sqrt{5 + \Zeff} \tau_\rmc \ln{\rmLambda_\rmc}}, \\
    s & = \frac{\sigma \EC}{j_0},
\end{align}
where \(\tau_\rmc = 4 \upi \varepsilon_0^2 m_\rme^2 c^3 / (n_\rme e^4 \lnLc ) \)  is the relativistic electron collision time, $m_\mathrm{e}$ is the electron mass, \(\varepsilon_0\) is the vacuum permittivity, \(\Zeff\) is the effective charge, \(E_\rmc = m_\rme c / (e \tau_\rmc) \) is the critical electric field introduced by \citet{Connor1975}, \(\lnLc = \ln{\rmLambda_\mathrm{th}} + 0.5\ln{(m_\rme c^2 /T)} \approx 14.6 + 0.5 \ln{(T_\mathrm{eV}/n_{\rme 20})}\) is the relativistic Coulomb logarithm, \(\ln{\rmLambda_\mathrm{th}} = 14.9      - 0.5\ln{n_{\rme 20}} + \ln{(T_\mathrm{eV}/1000)}\) is the thermal electron-electron Coulomb logarithm, \(T_\mathrm{eV}\) is the electron temperature in units of electronvolts, \(n_{\rme 20}\) is the electron density in units of \(10^{20}\: \mathrm{m}^{-3}\), and \(j_0 = I_{\mathrm{p}0}/A\) and \(I_\mathrm{p0}\) are the pre-disruption current density and current, respectively. The resulting dimensionless differential equations become
\begin{align}
    & \deriv{n}{t'} = \sum_\mathrm{seeds} \bar{\gamma}_\mathrm{seed} + \bar{\Gamma}_\mathrm{ava} n, \label{eq:normRunawayGeneration} \\
    & \deriv{ }{t'} \left( sE + n \right) = \frac{E_0 - E}{\alpha} \label{eq:normInductiveElectricField},
\end{align}
where,
\begin{align}
    E_0 & = \frac{E_{\parallel 0}}{\EC}, \\
    \alpha & = \frac{2}{\sqrt{5 + \Zeff}}\frac{L}{\mu_0 R_0} \frac{I_{\mathrm{p}0}}{I_\mathrm{A} \ln{\rmLambda_\rmc}}, \\
    \bar{\gamma}_\mathrm{seed} & = \sqrt{5 + \Zeff} \tau_\rmc \lnLc \frac{e c}{j_0} \gamma_\mathrm{seed}, \\
    \bar{\Gamma}_\mathrm{ava} & = \sqrt{5 + \Zeff} \tau_\rmc \lnLc \Gamma_\mathrm{ava}.
\end{align}
Here, \(I_\mathrm{A} = 4 \upi m_\mathrm{e} c /(\mu_0 e) \approx \SI{0.017}{\mega \ampere}\) is the Alfvén current. Note that using this normalization, \(n\) is equivalent to the conversion rate of the pre-disruption Ohmic current into a runaway electron current. Dividing equation \eqref{eq:normInductiveElectricField} by equation \eqref{eq:normRunawayGeneration}, the system is reduced to the single nonlinear ordinary differential equation
\begin{equation}
	s\deriv{E}{n} = -1 - \frac{E - E_0}{\alpha\left[\sum_\mathrm{seeds} \bar{\gamma}_\mathrm{seed} + \bar{\Gamma}_\mathrm{ava}n\right]}. \label{eq:nonlinearODE}
\end{equation}
Assuming that the runaway electron generation is  initially dominated by primary generation mechanisms, the avalanche source term can be neglected, leading to 
\begin{equation}
	s\deriv{E}{n} = -1 - \frac{E - E_0}{\alpha\sum_\mathrm{seeds} \bar{\gamma}_\mathrm{seed} }. \label{eq:nonlinearODEearly}
\end{equation}
At early times in the disruption, the electric field \(E\) exceeds \(E_0\) by a large margin. For example, in a high performance ITER plasma without material injection, \(E_0 \approx 0.05\), whereas \(E\) at the very least is larger than unity. In addition, \(\Sigma_\mathrm{seeds}\bar{\gamma}_\mathrm{seed} \) is assumed to be much smaller than \(E/\alpha\), i.e. the second term on the right hand side of equation (\ref{eq:nonlinearODEearly}) dominates. Using these assumptions, equation (\ref{eq:nonlinearODEearly}) reduces to
\begin{equation}
	s\deriv{E}{n} = - \frac{E}{ \alpha \sum_\mathrm{seeds} \bar{\gamma}_\mathrm{seeds} },
\end{equation}
which can be integrated to yield the seed density
\begin{equation}
    n_\mathrm{seed} = s \alpha \int_{\bar{E}_\rmc^\mathrm{eff}}^{E_1} \frac{\sum_\mathrm{seeds}\bar{\gamma}_\mathrm{seed}}{E}\,\rmd E.
\end{equation}
The upper integration limit is given by \(E_1\), which denotes the normalized electric field immediately after the thermal quench. The lower integration limit, \(\bar{E}_\rmc^\mathrm{eff}\), is the effective critical electric field \citep{Hesslow2018a} in units of \(\EC\).  \(\bar{E}_\rmc^\mathrm{eff}\) includes radiation forces and the enhanced collision frequency due to runaway electrons probing the internal structure of partially ionized ions.

The avalanche gain factor can, in principle, be estimated in a similar manner by assuming that after some short time (compared to the entire avalanche phase), the runaway electron density is sufficiently large for secondary generation to dominate, resulting in the avalanche gain factor
\begin{equation}
    n = n_\mathrm{seed} \exp({N_\mathrm{ava}}), \quad \quad N_\mathrm{ava} = s\alpha \int_{\bar{E}_\rmc^\mathrm{eff}}^{E_1} \frac{\bar{\Gamma}_\mathrm{ava}}{E} \, \rmd E.
\end{equation}
However, as the final runaway electron density is exponentially sensitive to \(N_\mathrm{ava}\), it is important to take into account the radial diffusion of the parallel electric field. This means that \(N_\mathrm{ava}\) has to be modified by a factor \(2 a_\mathrm{wall}^2/(x_1^2 a^2)\) \citep[see][]{Hesslow2019}, where \(a_\mathrm{wall}\) is the minor radius of the wall and \(x_1 \approx 2.4\) is the first zero of the zeroth order Bessel function of the first kind, \(J_0(x)\). \(N_\mathrm{ava}\) takes the form \citep{Hesslow2019}
\begin{equation}
    N_\mathrm{ava} = \tau_\mathrm{CQ} \int_{\bar{E}_\rmc^\mathrm{eff}}^{E_1} \frac{\Gamma_\mathrm{ava}}{E} \, \rmd E, \label{eq:semi-analyticalNava}
\end{equation}
where \(\tau_\mathrm{CQ}\) is the characteristic time scale of the current decay in the cylindrical limit and is given by 
\begin{equation}
    \tau_\mathrm{CQ} = \frac{L}{\mu_0 R_0}\frac{\sigma \mu_0 a_\mathrm{wall}^2}{x_1^2}.
\end{equation}
The exponent \(N_\mathrm{ava}\) is thus sensitive to the parameter \(a_\mathrm{wall}\), and the most accurate choice for \(a_\mathrm{wall}\) is not necessarily the minor radius of the first wall. Therefore, in this context, the \emph{wall} refers to the toroidally continuous conducting structure closest to the plasma, which is normally at a larger radius than the first wall of the vacuum chamber. The parameter \(a_\mathrm{wall}\) is made sufficiently large to contain a representative poloidal magnetic energy that is released during the disruption. Therefore, the values \(\ a_\mathrm{wall} = \SI{2.833}{\meter}\) and \( \SI{0.621}{\meter}\) are used in the ITER and SPARC simulations, respectively \citep{Pusztai2023,Ekmark2025}.

For an exponentially increasing runaway population, the assumption that the first term on the right-hand side of equation \eqref{eq:nonlinearODE} is negligible eventually becomes invalid, and the runaway electron current reaches unphysical values. However, the purpose of this model is not to determine the exact value of the runaway current, but rather to determine whether the exponential growth is sufficiently strong for the runaway current to become comparable to the pre-disruption Ohmic current.

\section{Avalanche generation and partial screening}
\label{sec:avalanche}
Runaway electron mitigation schemes usually involve massive material injection, where a mixture of deuterium and noble gases (primarily neon or argon) is injected into the plasma, either through gas puffing \citep{Reux2015,Pautasso2020} or in the form of cryogenic pellets \citep{Reux2022,Halldestam2025, Patel2025}. The main objective of the material injection is to enhance line radiation to reduce the localized thermal loads and electromagnetic forces on the device \citep{Bandyopadhyay2025}, as well as to increase the collisional drag, which leads to a larger effective critical electric field and critical momentum. Not only do free electrons contribute to the enhanced fraction force, but bound electrons as well, as energetic electrons may probe the internal structure of an ion, so that the nucleus is only partially screened by its bound electrons \citep{Hesslow2018a}. Accounting for this effect, the critical momentum becomes \citep{DREAM}
\begin{equation}
    p_\rmc^2 = \frac{\sqrt{4 \nubar{s}(p_\star)^2 + \nubar{s}(p_\star) \nubar{D}(p_\star) }}{E - \bar{E}_\rmc^\mathrm{eff}},
\end{equation}
where \(\bar{E}_\rmc^\mathrm{eff}\) is evaluated using expressions (23) and (24) in \citet{Hesslow2018a} and \(\nubar{D}\) and \(\nubar{s}\) denote the normalized deflection and slowing-down frequencies, respectively, and are defined as \citep{Hesslow2018b}
\begin{align}
    \nubar{D}(p) & = 1 + \Zeff \frac{\ln{\rmLambda^{\rme \mathrm{i}}}}{\ln{\rmLambda^{\rme\rme}}} \sum_j \frac{n_j}{n_\rme} g_j(p), \\
    \nubar{s}(p) & = 1 + \frac{1}{\ln{\rmLambda^{\rme\rme}}}\sum_j \frac{n_j}{n_\rme} N_{\rme,j} \left[ \frac{1}{k} \ln{(1 + h_j^k)} - \beta^2 \right],
\end{align}
where the functions \(g_j\) and \(h_j\) are defined as
\begin{align}
    g_j(p) & = \frac{2}{3} \left( Z_j^2 - Z_{0,j}^2 \right) \ln{\left[(p \bar{a}_j)^{3/2} + 1\right]} - \frac{2}{3} \frac{N_{\rme,j}^2 (p \bar{a}_j)^{3/2}}{(p \bar{a}_j)^{3/2} + 1}, \\
    h_j(p) & = p\sqrt{\gamma - 1} \frac{m_\rme c^2}{I_j}.
\end{align}
In the expressions above, \(Z_j\) denotes the atomic number of ion species \(j\), \(Z_{0,j}\) denotes the charge number, and \(N_{\rme,j} = Z_j - Z_{0,j}\) is the number of bound electrons. Furthermore, \(p\) is the electron momentum in units of \(m_\rme c\), \(\beta\) is the electron velocity in units of \(c\), and \(\gamma\) is the Lorentz factor. The atomic data \(\bar{a}_j\) and \(I_j\) denote the normalized effective length scale and the mean excitation energy, respectively, and are specific to each ion and charge state. They have been tabulated by \citet{Hesslow2018b} and \citet{Sauer2018}, respectively. The ad-hoc parameter \(k = 5\) is an interpolation parameter between the low and high energy regimes. The electron-electron and electron-ion Coulomb logarithms, \(\ln{\rmLambda^{\rme \rme}}\) and \(\ln{\rmLambda^{\rme \mathrm{i}}}\), are evaluated as
\begin{align}
    \ln{\rmLambda^{\rme \rme}} & = \ln{\rmLambda_\mathrm{th}} + \frac{1}{k}\ln{\left[ 1 + \left(\frac{2(\gamma - 1)}{p_\mathrm{th}^2}\right)^{k/2} \right]},\\
    \ln{\rmLambda^{\rme \mathrm{i}}} & = \ln{\rmLambda_\mathrm{th}} + \frac{1}{k}\ln{\left[ 1 + \left(\frac{2p}{p_\mathrm{th}} \right)^k\right]},
\end{align}
where \(p_\mathrm{th} = \sqrt{2 T_\rme/(m_\rme c^2)}\) is the normalized thermal momentum. The normalized slowing-down and deflection frequencies are evaluated at \(p = p_\star\), which is defined implicitly as
\begin{equation}
    p_\star^2 = \frac{\sqrt{\nubar{s}(p_\star) \nubar{D}(p_\star)}}{E},
\end{equation}
and has to be evaluated numerically. In a fully ionized plasma (or in the low energy limit), \(\nubar{s} \rightarrow 1\) and \(\nubar{D} \rightarrow 1 + \Zeff\). Implementing the model for partially screened ion nuclei derived by \citet{Hesslow2018b} would make the otherwise analytical model a semi-analytical model, in the sense that some simple numerical evaluations would be required. For many applications, this is not an obstacle. For this reason, two versions of the criterion will be presented in \S\ref{sec:criterion}, namely an analytical criterion that consists solely of analytical expressions, and a semi-analytical criterion of a similar nature but with a more detailed form of the seed densities and the avalanche gain factor. In the analytical model, the critical momentum is evaluated in its completely screened limit and reduces to \(p_\rmc^2 = \sqrt{5 + \Zeff}/(E - 1) \).

Massive material injection not only increases collisional drag but also provides more target electrons during the avalanche, which could enhance the avalanche gain factor by many orders of magnitude if the composition of the injected material is chosen poorly. A model that captures this behaviour was developed by \citet{Hesslow2019} and later modified by \citet{DREAM} to improve the accuracy of the avalanche generation rate in nearly neutral low-\(Z\) plasmas. The avalanche generation rate is given by
\begin{equation}
    \Gamma_\mathrm{ava} = \frac{1}{\tau_\rmc \lnLc}\frac{n_\rme^\mathrm{tot}}{n_\rme} \frac{E - \bar{E}_\rmc^\mathrm{eff}}{\sqrt{4 \nubar{s}(p_\star)^2 + \nubar{s}(p_\star) \nubar{D}(p_\star) }}, \label{eq:GammaPS}
\end{equation}
where \(n_\rme^\mathrm{tot}\) is the total number of electrons (free and bound) in the plasma. In the limit of a fully ionized plasma, the avalanche generation rate derived by \citet{Rosenbluth1997} is recovered. To allow for analytical treatment, the  avalanche generation rate of Rosenbluth-Putvinski can be adjusted slightly to approximate the effect of partially ionized ions by including the bound target electrons and by modifying the critical electric field \citep{Putvinski1997} and the effective charge \citep{Parks1999}. To be consistent with the notation in \citet{Hesslow2018b}, these will be denoted \(E_\rmc^\mathrm{RP}\) and \(\Zeff^\mathrm{RP}\), respectively. The approximate avalanche generation rate becomes
\begin{equation}
    \Gamma_\mathrm{ava}^\mathrm{RP} = \frac{1}{\tau_\rmc \lnLc}\frac{n_\rme^\mathrm{tot}}{n_\rme} \frac{E/\bar{E}_\rmc^\mathrm{RP} - 1}{\sqrt{5 + \Zeff^\mathrm{RP} }},
\end{equation}
where
\begin{align}
    E_\rmc^\mathrm{RP} & = \frac{1}{2}\left( 1 + \frac{n_\rme^\mathrm{tot}}{n_\rme}\right)E_\rmc \equiv \bar{E}_\rmc^\mathrm{RP} E_\rmc, \\
    \Zeff^\mathrm{RP} & = \sum_{\substack{j \: \text{part.} \\ \text{ionized}}} \frac{n_j}{n_\rme} \frac{Z_j^2}{2} + \sum_{\substack{j \: \text{fully} \\ \text{ionized}}} \frac{n_j}{n_\rme} Z_j^2.
\end{align}
The exponent in the avalanche gain factor becomes
\begin{equation}
    N_\mathrm{ava}^\mathrm{RP} = \frac{1}{\lnLc \sqrt{5 + \Zeff^\mathrm{RP} }} \frac{\tau_\mathrm{CQ}}{\tau_\rmc }\frac{n_\rme^\mathrm{tot}}{n_\rme} \left[ \frac{E_1}{\bar{E}_\rmc^\mathrm{RP}} - \ln{\left(\frac{E_1}{\bar{E}_\rmc^\mathrm{RP}}\right)} - 1\right] \Heavi{\left(E_1 - \bar{E}_\rmc^\mathrm{RP}\right)}, \label{eq:analyticalNava}
\end{equation}
where the Heaviside function \(\Heavi\) reflects that avalanche generation occurs only when \(E_1 > \bar{E}_\rmc^\mathrm{RP}\).

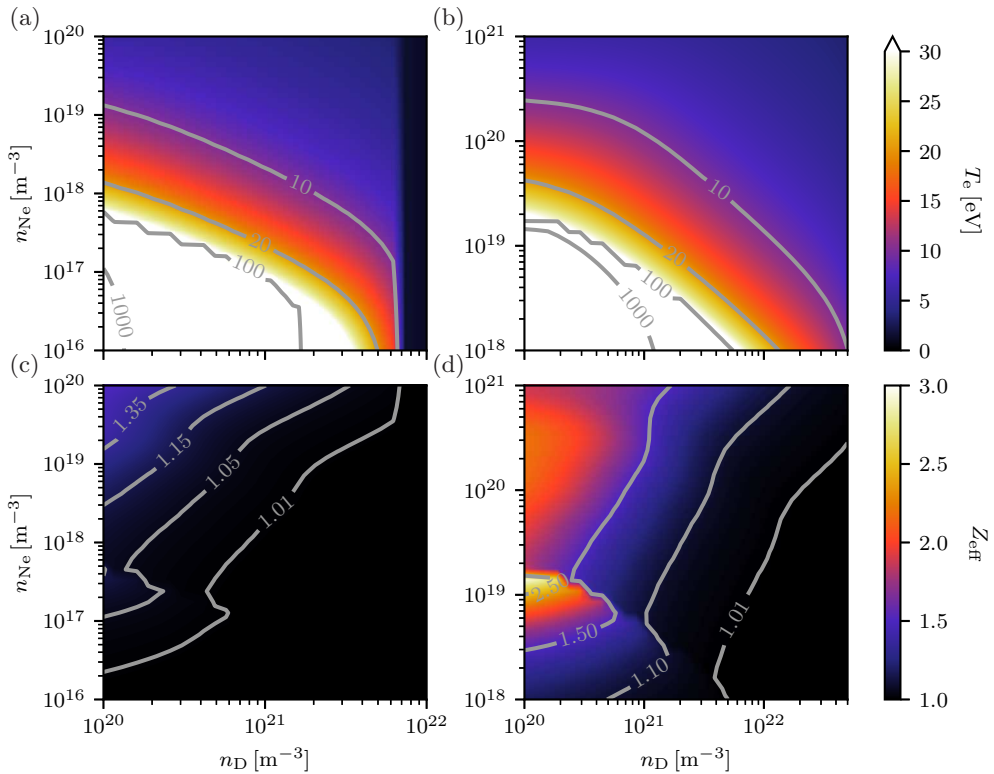
\begin{figure}
    \centering
    \includegraphics{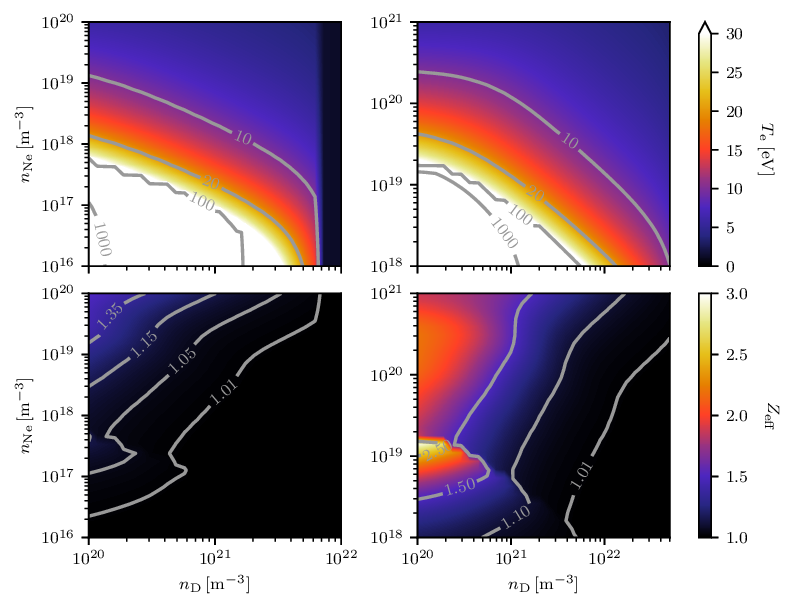}
    \put(-376,286){(a)}
    \put(-217,286){(b)}
    \put(-376,156){(c)}
    \put(-217,156){(d)}
    \caption{Equilibrium electron temperature evaluated using equations \eqref{eq:OhmicRadiativeBalance}-\eqref{eq:chargeStateDistribution} for (a) ITER and (b) SPARC. The effective charge \(\Zeff\) evaluated using a charge state distribution consistent with the temperature for (c) ITER and (d) SPARC. The quantities on the axes denote injected deuterium and neon densities. As the colour scale of \(T_\rme\) is saturated to highlight interesting ranges, additional contours are added outside these ranges. Note also, that in particular the neon concentration ranges are different in ITER and SPARC.}
    \label{fig:T_e}
\end{figure}

In this work, the ion charge state distribution and electron temperature are determined assuming that the plasma is in a collisional-radiative equilibrium where the Ohmic heating is balanced by the line radiation and bremsstrahlung \citep[see][]{MartinSolis2017,Vallhagen2020} according to
\begin{align}
    \frac{j_0^2}{\sigma(T_\rme,Z_\mathrm{eff})} & = \sum_{i,l} n_\rme n_i^l L_i^l (T_\rme,n_\rme), \label{eq:OhmicRadiativeBalance} \\
    n_i^{l} & = n_i\left(
			1+
			\sum_{j=0}^{l-1}\prod_{k=j+1}^{l}\frac{R_i^{k}}{I_i^{k-1}} +
			\sum_{j=l+1}^Z\prod_{k=l}^{j-1}\frac{I_i^{k}}{R_i^{k+1}}
		\right)^{-1}, \label{eq:chargeStateDistribution}
\end{align}
where \(n_i^l\) is the density of ion species \(i\) at charge state \(l\), \(L_i^l(T_\rme, n_\rme)\) is the combined rate coefficient for bremsstrahlung and line radiation from excitation and recombination, \(I_i^{l} = I_i^{l}(T_\rme,n_\rme)\) denotes the ionization rate of ion species \(i\) from charge state \(l\) to charge state \(l + 1\), \(R_i^{l} = R_i^{l}(T_\rme,n_\rme)\) denotes the recombination rate of ion species \(i\) from charge state \(l\) to charge state \(l - 1\), and \(n_i\) is the total density of ion species \(i\). The ion charge state distribution (and thereby \(n_\rme\) and \(\Zeff\)) and the electron temperature are initially evaluated self-consistently for the given plasma composition and pre-disruption current density. The temperature is then kept constant. Ionization, recombination, and radiation rates are obtained from the Atomic Data and Analysis Structure (ADAS) database\footnote{OPEN-ADAS database: \url{https://open.adas.ac.uk/}}. Determining the temperature from a collisional-radiative equilibrium --- along with neglecting time-dependent impurity sources --- implies that the detailed time evolution of the atomic physics is not accounted for, which can affect the accuracy of both the temperature evolution and the runaway electron generation process \citep{Vallhagen2020}. This approximation is nevertheless required to maintain the semi-analytical tractability of the problem, as it avoids explicit time integration of the rate equations and provides a level of accuracy sufficient for the purposes of the model.

Equilibrium temperatures and the associated effective charge \(\Zeff\) for a range of deuterium and neon concentrations are shown in figure \ref{fig:T_e}. The current density is set to \(j_0 = I_\mathrm{p0}/(\upi a^2) \), which becomes \(\SI{1.19}{\mega \ampere \per \meter \squared}\) and \(\SI{10.0}{\mega \ampere \per \meter \squared}\) in ITER and SPARC, respectively. \(\Zeff\) can be used as a measure of the degree of ionization of the injected neon and is shown in figures \ref{fig:T_e}c and \ref{fig:T_e}d. Although neon can have a large degree of ionization in the hot lower left quadrant, it is too diluted to contribute substantially to \(\Zeff\). This is primarily true for ITER, where neon densities down to \SI{1e16}{\per \meter \cubed} are considered. As the neon concentration increases, more power is radiated away, and the temperature drops, leading to a lower degree of ionization. In SPARC (figure \ref{fig:T_e}d), \(\Zeff\) peaks around \(n_\mathrm{Ne} = \SI{1e19}{\per \meter \cubed}\) at \(\Zeff \approx 2.9\), and this trade-off between plasma temperature and relative neon concentration is clearly visible. In ITER (figure \ref{fig:T_e}c), however, due to the lower current density, the temperature is lower, and \(\Zeff\) peaks in the top left corner at a value of \(\Zeff \approx 1.5\).

\section{Nuclear sources}
\label{sec:nuclearsources}
In activated devices, seed runaway electrons can be generated from nuclear sources such as tritium beta decay and Compton scattering of gamma photons originating from the irradiated plasma-facing components. These sources can be significant, in the sense that both sources can (under certain circumstances, see \S \ref{sec:criterion}) produce enough seed electrons for them to avalanche into mega-ampere runaway electron currents. In this section, the nuclear seed generation rates are presented, and it is shown how they can be integrated to be compatible with an analytical criterion for significant runaway electron generation.

\subsection{Tritium decay}
\begin{figure}
  \centering
  \includegraphics{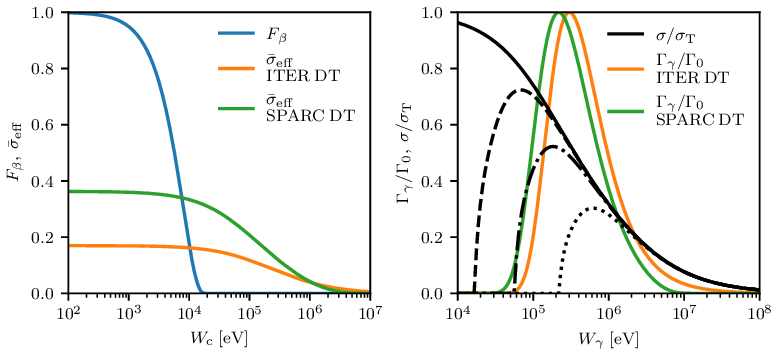}
  \put(-375,168){(a)}
  \put(-193,168){(b)}
  \caption{Fraction of \(\beta\)-electrons born in the runaway regime, \(F_\beta\), (blue) and normalized Compton cross-section averaged over the \(\gamma\)-spectrum, \(\bar{\sigma}_\mathrm{eff}\), plotted against critical energy \(W_\rmc\) (a). The effective cross-section is plotted for DT plasmas in both ITER (orange) and SPARC (green), and is obtained by averaging the cross-section (black) over the corresponding Compton spectrum \(\Gamma_\gamma\) (b). The cross-section is plotted for critical energies  0, 1, 10, and \SI{100}{\kilo \electronvolt} (solid, dashed, dot-dashed, and dotted curves, respectively).}
  \label{fig:sigmaEffFbeta}
\end{figure}

Tritium has a half-life of \(\tau_\mathrm{T} = \SI[separate-uncertainty = true]{4500(8)}{\day} \) and decays as
\begin{equation}
    \mathrm{T} \rightarrow {}^3_2\mathrm{He} + \rme^- + \nubar{e}+ \SI{18.6}{\kilo \electronvolt}.
\end{equation}
On average, the electron receives an energy of \SI{5.7}{\kilo \electronvolt}, but all energies from 0 to \(W_\mathrm{max} = \SI{18.6}{\kilo \electronvolt}\) are possible. The production rate of beta electrons is given by 
\begin{equation}
    \deriv{n_\beta}{t} = \ln{2} \frac{n_\mathrm{T}}{\tau_\mathrm{T}},
\end{equation}
where \(n_\mathrm{T}\) is the tritium density. In a post-disruption plasma with temperatures up to \SI{100}{\electronvolt}, beta electrons can be considered relatively energetic, and some fraction \(F_\beta(W_\rmc)\) of these will be born above the critical energy \(W_\rmc = m_\rme c^2 (\sqrt{p_\rmc^2 + 1} - 1 )\). The runaway electron generation due to tritium decay is then given by \citep{MartinSolis2017,Fulop2020}
\begin{equation}
    \gamma_\mathrm{T}(W_\rmc) = \ln{2} \frac{n_\mathrm{T}}{\tau_\mathrm{T}} F_\beta(W_\rmc),
\end{equation}
where \(F_\beta(W_\rmc)\) is given by
\begin{equation}
    F_\beta(W_\rmc) = \frac{\int_{W_\rmc}^{W_\mathrm{max}} F(p,2)p\mathcal{W}(W_\mathrm{max} - W)^2\rmd W}{\int_{0}^{W_\mathrm{max}} F(p,2)p\mathcal{W}(W_\mathrm{max} - W)^2\rmd W},
    \label{eq:full_fbeta}
\end{equation}
\(F(p,Z)\) is the Fermi function, and \(\mathcal{W} = m_\rme c^2 \gamma\) is the total energy of the electron. Neglecting the interaction between the positive nucleus and the beta electron, the fraction of beta electrons born above the critical energy is given by \citep{MartinAndShaw2019,Fulop2020}
\begin{align}
    F_\beta(W_\rmc) & \approx \frac{\int_{W_\rmc}^{W_\mathrm{max}} p(W_\mathrm{max} - W)^2\rmd W}{\int_{0}^{W_\mathrm{max}} p(W_\mathrm{max} - W)^2\rmd W} \Heavi{(W_\rmc-W_\mathrm{max})}\\
    & = \left(1 - \frac{35}{8}  w^{3/2} + \frac{21}{4} w^{5/2} - \frac{15}{8} w^{7/2}\right) \Heavi{(w-1)},
    \label{eq:approx_fbeta}
\end{align}
where \(w = W_\rmc/W_\mathrm{max}\). \(F_\beta(W_\rmc)\) is plotted against the critical energy in figure~\ref{fig:sigmaEffFbeta}a.

If the critical energy is larger than \(W_\mathrm{max}\), no runaway electrons are generated according to this fluid model. Note that in a kinetic model, energetic electrons born below the critical energy can end up in the runaway regime through collisional processes \citep{Ekmark2024}. Using the same normalization as in \S\ref{sec:fluidEquations}, the dimensionless runaway electron generation rate due to tritium beta decay takes the form
\begin{equation}
    \bar{\gamma}_\mathrm{T} = \ln{2}\sqrt{5 + \Zeff} \lnLc \frac{ec n_\mathrm{T}}{j_0} \frac{\tau_\rmc}{\tau_\mathrm{T}}F_\beta(W_\rmc),
\end{equation}
and the tritium seed density becomes
\begin{equation}
    n_\mathrm{seed,T} = \frac{x_1^2 \ln{2}}{2} \left( \frac{a}{a_\mathrm{wall}} \right)^2 \frac{ec n_\mathrm{T}}{j_0} \frac{\tau_\mathrm{CQ}}{\tau_\mathrm{T}} \int_{E_\mathrm{min}}^{E_1}  \frac{F_\beta(W_\rmc)}{E} \, \rmd E,\label{eq:semi-analyticalTseed}
\end{equation}
where \(E_\mathrm{min}\) is the value of the normalized electric field \(E\) that corresponds to \(W_\rmc = W_\mathrm{max}\). Assuming complete screening of the ion nuclei, \(E_\mathrm{min}\) becomes
\begin{equation}
    E_\mathrm{min} =  \frac{\sqrt{5 + \Zeff}}{w_\mathrm{max}^2 + 2 w_\mathrm{max}} + 1 \approx 34,
\end{equation}
where \(w_\mathrm{max} = W_\mathrm{max}/(m_\rme c^2)\) is the maximum beta decay energy normalized to the electron rest energy, and the numerical evaluation assumes that \(\Zeff = 1\).

In the limit of complete screening of the ion nuclei, \(F_\beta(W_\rmc)/E\) can be integrated analytically to become
\begin{equation}
\begin{split}
    \int \frac{F_\beta(W_\rmc)}{E} \, \rmd E = \ln{E} - \sqrt{5+\Zeff}w_\mathrm{max}^{-7/2} w_\rmc^{1/2} \left(21 w_\mathrm{max} + \frac{45}{2} - \frac{5}{2} w_\rmc\right) \\
     + \sum_{i = 1}^3 c_n \left(\frac{1}{w_\mathrm{max}}\right)^{(2n + 1)/2} G_{2n + 1}\left(\Zeff;w_\rmc\right), \label{eq:analyticalTseed}
\end{split}
\end{equation}
where \(w_\rmc = W_\rmc/(m_\rme c^2)\) is the critical energy normalized to the electron rest energy, and the coefficients \(c_1\) through \(c_3\) are given by 35/4, 21/2, and 15/4, respectively. The function \(G_{n}\left(\Zeff;w_\rmc\right)\) is defined as
\begin{align}
     G_{n}\left(\Zeff;w_\rmc\right) & = 2^{n/2}\arctan{\left(  \sqrt{\frac{w_\rmc}{2}}\right)} \nonumber\\ & \quad- (5 + \Zeff)^{n/8} T_n(u) \arctan{\left(\frac{2 (5 + \Zeff)^{1/8} u \sqrt{w_\rmc} }{(5 + \Zeff)^{1/4} - w_\rmc}\right)} \\
    & \quad- (5 + \Zeff)^{n/8} v U_{n - 1} (u) \arctanh{\left(\frac{2 (5 + \Zeff)^{1/8} v \sqrt{w_\rmc}}{(5 + \Zeff)^{1/4} + w_\rmc}\right)} \nonumber,
\end{align}
where \(T_n\) and \(U_n\) are Chebyshev polynomials of the first and second kind, respectively, and
\begin{align}
    u & = \frac{1}{\sqrt{2}} \sqrt{1 + (5 + \Zeff)^{-1/4}}, \\
    v & = \frac{1}{\sqrt{2}} \sqrt{1 - (5 + \Zeff)^{-1/4}}.
\end{align}

\subsection{Compton scattering}

The activated wall radiates gamma photons up to the MeV range that can, through Compton scattering, launch both free and bound electrons into the runaway region of momentum space. In this work, the energy spectrum \(\Gamma_\gamma (W_\gamma)\) is given as a closed-form expression by fitting the function
\begin{align}
    \Gamma_\gamma (W_\gamma) & = \Gamma_0 \exp{[-\exp{(-z)} - z + 1]}, \\
    z & = \frac{\ln{W_\gamma\,[\mathrm{MeV}]} + C_1}{C_2} + C_3(W_\gamma\, [\mathrm{MeV}])^2,
\end{align}
to radiation transport calculations. The gamma spectrum is machine- and shot-dependent, and fitting parameters \(C_i\) for a few relevant machines and plasma compositions \citep{MartinSolis2017,Ekmark2025} are reproduced in table \ref{tab:Gamma}. The normalization constant \(\Gamma_0\) ensures that the total photon flux becomes \(\Gamma_\mathrm{flux}\).

The rate at which electrons are scattered into the runaway region by gamma photons emitted with a certain energy \(W_\gamma\) is proportional to the cross-section \(\sigma(W_\gamma,W_\rmc)\), which is obtained by integrating the Klein-Nishina differential cross-section \citep{KleinNishina}
\begin{equation}
    \deriv{\sigma}{\Omega} = \frac{r_\rme^2}{2} \frac{W_\gamma'^2}{W_\gamma^2} \left( \frac{W_\gamma}{W_\gamma'} + \frac{W_\gamma'}{W_\gamma} - \sin^2{\theta}\right),
\end{equation}
where \(r_\rme = e^2/(4\upi\varepsilon_0 m_\rme c^2)\) is the classical electron radius, and \(\theta\) and \(W_\gamma'\) are the deflection angle and the energy of the scattered photon, respectively. \(W_\gamma'\) can be related to the incident photon energy and the deflection angle through the kinematic relation
\begin{equation}
    W_\gamma' = \frac{W_\gamma}{1 + \frac{W_\gamma}{m_\rme c^2} (1 - \cos{\theta})}.
\end{equation}
By letting \(W_\gamma-W_\gamma' = W_\rmc\), the kinematic relation above can be used to find the critical angle \(\theta_\rmc\), i.e. the minimum photon deflection angle required to scatter an electron into the runaway region for a given photon energy and critical energy. The critical angle becomes
\begin{equation}
    \cos{\theta_\rmc} = 1 - \frac{w_\rmc}{w_\gamma} \frac{1}{w_\gamma - w_\rmc},
\end{equation}
yielding the integrated Compton cross-section
\begin{eqnarray}
    \sigma(W_\gamma,W_\rmc) & = & 2\upi\int_{\theta_\rmc}^\upi \deriv{\sigma}{\Omega} \sin{\theta} \, \rmd \theta \nonumber \\
    & = &\frac{3\sigma_\mathrm{T}}{8} \Bigg\{ \frac{w_\gamma^2 -2w_\gamma -2}{w_\gamma^3} \ln{\left[\frac{1 + 2w_\gamma}{1 + w_\gamma(1 - \cos{\theta_\rmc})}\right]} \nonumber \\
     &&  + \frac{1}{2w_\gamma} \left[ \frac{1}{\left[ 1 + w_\gamma (1 - \cos{\theta_\rmc}) \right]^2} - \frac{1}{(1 + 2w_\gamma)^2} \right] \\
     && - \frac{1}{w_\gamma^3}\left[ 1 - w_\gamma -\frac{1 + 2w_\gamma}{1 + w_\gamma(1-\cos{\theta_\rmc})} - w_\gamma\cos{\theta_\rmc}\right]\Bigg\} \nonumber,
\end{eqnarray}
where \(\sigma_\mathrm{T} = 8 \upi r_\rme^2/3\) is the Thomson scattering cross-section and \(w_\gamma = W_\gamma/(m_\rme c^2)\) is the photon energy normalized to the electron rest energy \citep{MartinSolis2017}.

\begin{table}   
    \setlength{\tabcolsep}{5pt}
    \begin{center}
    \def~{\hphantom{0}}
    \begin{tabular}{lcccccc}
        & \(C_1\) & \(C_2\) & \(C_3\) & \(\Gamma_\mathrm{0}\) [m\(^{-2}\)s\(^{-1}\)eV\(^{-1}\)] & \(\Gamma_\mathrm{flux}\) [m\(^{-2}\)s\(^{-1}\)] & \(\bar{\sigma}_\mathrm{eff}(0)\)\\[3pt]
        ITER DT & 1.2 & 0.8 & 0 & \(3.3 \times 10^{11}\) &\(1.0 \times 10^{18}\) & \(0.17 \) \\
        SPARC DD & 1.627 & 0.919 & 0.094 & \(3.304 \times 10^{9}\) & \(3.3 \times 10^{15}\) & \(0.3921 \) \\
        SPARC DT & 1.525 & 0.850 & 0.038 & \(1.254 \times 10^{12}\) &\(1.4 \times 10^{18}\) & \(0.3630 \) \\
    \end{tabular}
    \caption{Fitting coefficients for gamma photon spectra for three reference scenarios \citep{MartinSolis2017,Ekmark2025}. The final column explicitly lists the value of \(\bar{\sigma}_\mathrm{eff}(0)\) used in equation \eqref{eq:analyticalCseed}.}
    \label{tab:Gamma}
    \end{center}
\end{table}

Using the gamma spectrum and cross-section above, the runaway electron generation rate due to Compton scattering can be approximately evaluated as
\begin{equation}
    \gamma_\mathrm{C}(W_\rmc) \approx n_\rme^\mathrm{tot} \int_0^\infty \Gamma_\gamma (W_\gamma) \sigma(W_\gamma,W_\rmc) \rmd W_\gamma \equiv n_\rme^\mathrm{tot} \Gamma_\mathrm{flux} \sigma_\mathrm{eff}(W_\rmc), \label{eq:comptonRate}
\end{equation}
where \(\sigma_\mathrm{eff}(W_\rmc)\) is the Compton cross-section averaged over the energy spectrum. The normalized seed density becomes
\begin{equation}
    n_\mathrm{seed,C} = \frac{x_1^2}{2} \left( \frac{a}{a_\mathrm{wall}} \right)^2 \frac{e c n_\rme^\mathrm{tot}}{j_0}  \tau_\mathrm{CQ} \Gamma_\mathrm{flux} \sigma_\mathrm{T} \int_{\EceffNorm}^{E_1} \frac{\bar{\sigma}_\mathrm{eff}(W_\rmc)}{E} \, \rmd E, \label{eq:semi-analyticalCseed}
\end{equation}
where \(\bar{\sigma}_\mathrm{eff} \equiv \sigma_\mathrm{eff}/\sigma_\mathrm{T}\) is plotted against the critical energy in figure \ref{fig:sigmaEffFbeta}a for DT plasmas in both ITER and SPARC. An upper bound on the Compton seed can be established by assuming that all electrons struck by a gamma photon run away. This overestimates the Compton seed by approximately half an order of magnitude compared to expression \eqref{eq:semi-analyticalCseed}, so a reasonable estimate of the integrated Compton seed becomes
\begin{equation}
    n_\mathrm{seed,C} \approx \frac{x_1^2}{4} \left( \frac{a}{a_\mathrm{wall}} \right)^2 \frac{e c n_\rme^\mathrm{tot}}{j_0}  \tau_\mathrm{CQ} \Gamma_\mathrm{flux} \sigma_\mathrm{T} \bar{\sigma}_\mathrm{eff}(0) \ln{\left(\frac{E_1}{\EceffNorm}\right)}, \label{eq:analyticalCseed}
\end{equation}
where a factor of 1/2 has been included to compensate for (part of) the overestimation. For convenience, the value of \(\bar{\sigma}_\mathrm{eff}(0)\) is given in table \ref{tab:Gamma} for each Compton spectrum.

As the plasma disrupts and the fusion reactions cease, the energy spectrum is assumed to retain its shape but have a photon flux that is three to four orders of magnitude lower than the fluxes listed in table \ref{tab:Gamma} \citep{MartinSolis2017,Vallhagen2024,Ekmark2025}. The model presented in this work has no explicit temporal resolution and is therefore unable to include a time-dependent photon flux; thus, the photon flux is set to be a constant factor of 1000 smaller than the value listed in table \ref{tab:Gamma}. The criterion is only logarithmically sensitive to the seed densities, so using the exact value is not essential to the validity of the model. In addition, early seed electrons produced before the fusion reactions cease are prone to radial losses, increasing the relevance of the reduced photon flux.    

\section{Criterion for significant runaway electron generation}
\label{sec:criterion}
The runaway electron generation is assumed to be significant if \(n \sim 1 \), or equivalently
\begin{equation}
    \mathcal{Z} \equiv N_\mathrm{ava} + \ln{n_\mathrm{seed}} > 0, \label{eq:criterion}
\end{equation}
where, again, \(n\) is the normalized runaway electron density introduced in \S\ref{sec:fluidEquations}. Note that \(n\) is equivalent to the ratio between the runaway electron current and the pre-disruption current, and that \(n\) is allowed to be larger than unity, as the saturation of the runaway electron current was neglected in \S\ref{sec:fluidEquations}. This criterion can be used to assess whether a post-disruption plasma with a given set of parameters can be expected to produce a large number of runaway electrons. \(\mathcal{Z}\) can be evaluated using either the analytical expressions \eqref{eq:analyticalNava}, \eqref{eq:analyticalTseed}, and \eqref{eq:analyticalCseed}, or the semi-analytical expressions \eqref{eq:semi-analyticalNava}, \eqref{eq:semi-analyticalTseed}, and \eqref{eq:semi-analyticalCseed}\footnote{A numerical implementation of both the analytical and semi-analytical formulations is available at \url{https://github.com/chalmersplasmatheory/Zcriterion.git}}. This makes the criterion a useful tool for quickly determining favourable regions in parameter space. To verify the criterion, it is compared to zero-dimensional fluid simulations using the \textsc{Dream} code \citep{DREAM} for both ITER- and SPARC-relevant parameters. Note that only tritium and Compton seeds are considered, as similar criteria based on Dreicer and hot-tail generation have previously been studied elsewhere \citep[see][]{Helander2002,Fulop2009}.

\begin{figure}
  \centering
  \includegraphics{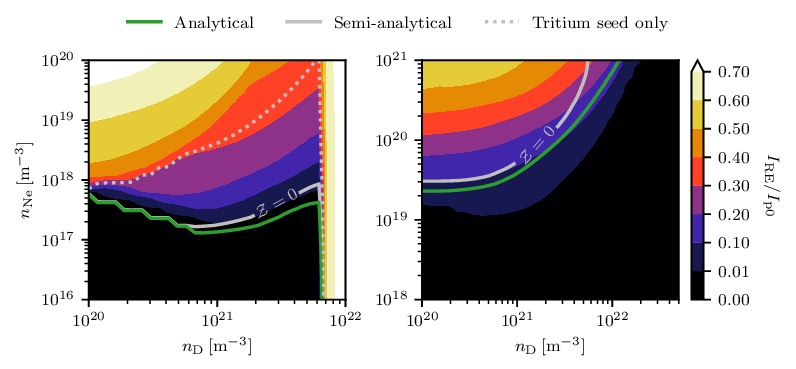}
  \put(-375,153){(a)}
  \put(-215,153){(b)}
  \caption{\textsc{Dream} simulations in 0D (filled contours) compared to inequality \eqref{eq:criterion} (green and grey contours for analytical and semi-analytical expressions, respectively) using (a) ITER-relevant and (b) SPARC-relevant parameters. The criterion \(\mathcal{Z} = 0\) approximately delineates regions in parameter space where a significant fraction of the Ohmic current is converted into a runaway electron current. The grey dotted contour represents \(\mathcal{Z} = 0\), evaluated using only the tritium seed. In the SPARC case, the dotted contour overlaps with the solid contour. Note the nonlinearity in the lower part of the colour map.}
  \label{fig:criterion}
\end{figure}
The result of this comparison is illustrated in figure \ref{fig:criterion}, where the filled contours represent the fraction of the Ohmic current converted into runaway electron current over a large range of injected deuterium and neon densities. The initial plasma current was set to \(I_{\mathrm{p}0} = \SI{15}{\mega \ampere}\) in ITER and \(I_{\mathrm{p}0} = \SI{8.7}{\mega \ampere}\) in SPARC, and the tritium density (as well as the initial deuterium density) was set to \(n_\mathrm{T} = \SI{5e19}{\per \meter \cubed}\) in ITER and \(n_\mathrm{T} = \SI{2e20}{\per \meter \cubed}\) in SPARC. The temperature was taken to be constant throughout the simulation and was determined by solving equations \eqref{eq:OhmicRadiativeBalance}-\eqref{eq:chargeStateDistribution} for \(T_\mathrm{e}\) for each combination of \(n_\mathrm{D}\) and \(n_\mathrm{Ne}\), using the pre-disruption Ohmic current.

As shown in figure \ref{fig:criterion}, the contour \(\mathcal{Z} = 0\) approximately delineates the region in parameter space where the zero-dimensional fluid model in \textsc{Dream} predicts 1\% runaway electron current conversion in ITER and 10-30\%  runaway electron current conversion in SPARC. A runaway electron current conversion of 1\% in ITER is consistent with the requirement to keep the runaway electron current below \(\SI{150}{\kilo \ampere}\) \citep{Lehnen2021Talk}, but the criterion is less conservative for SPARC-like parameters due to the exponential dependency on the initial plasma current. However, it is possible to compensate for this by explicitly making the criterion more conservative, i.e. defining significant runaway electron generation as e.g. \(n\sim0.1\) or \(n\sim 0.01\).

\begin{figure}
  \centering
  \includegraphics{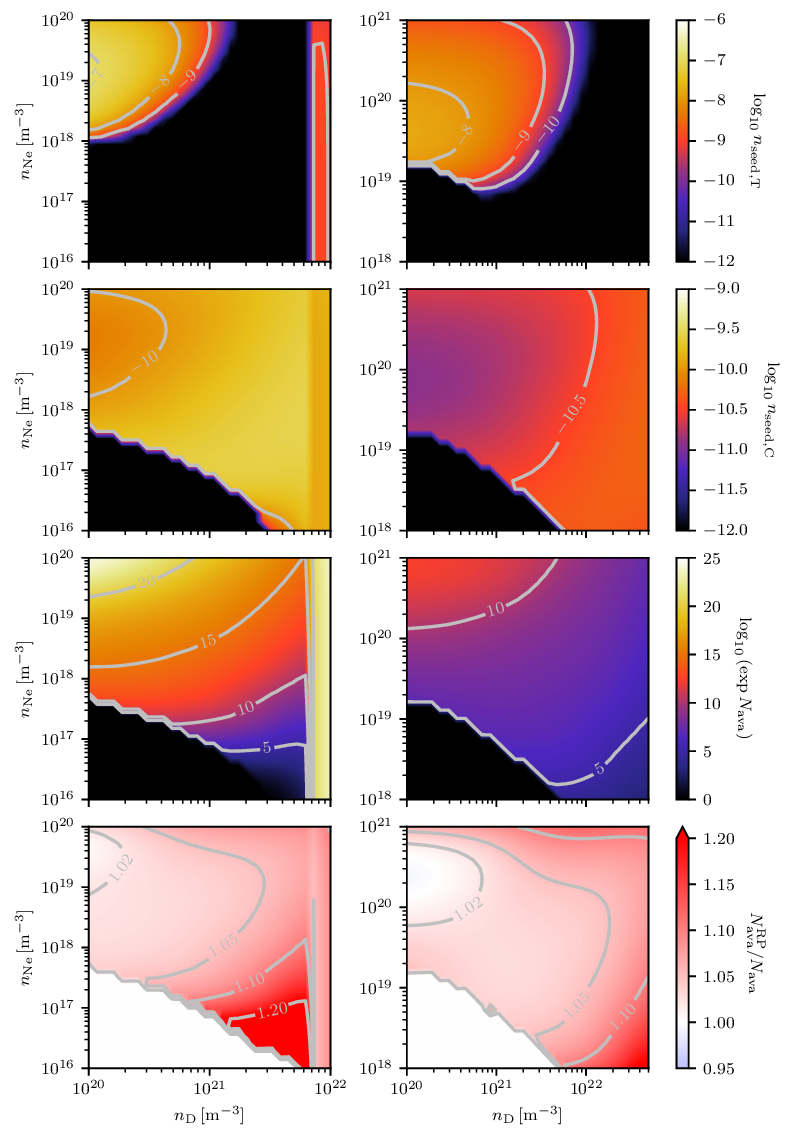}
  \put(-376,544){(a)}
  \put(-222,544){(b)}
  \put(-376,414){(c)}
  \put(-222,414){(d)}
  \put(-376,284){(e)}
  \put(-222,284){(f)}
  \put(-376,154){(g)}
  \put(-222,154){(h)}
  \caption{Tritium (a and b) and Compton (c and d) seeds evaluated using equations \eqref{eq:semi-analyticalTseed} and \eqref{eq:semi-analyticalCseed}, respectively. The avalanche gain factor (e and f) is evaluated using the semi-analytical expression \eqref{eq:semi-analyticalNava}. Furthermore, the semi-analytical formulation is compared to the analytical expression \eqref{eq:analyticalNava} (g and h). The expressions are evaluated for ITER-like parameters in the left column, and SPARC-like parameters in the right column.}
  \label{fig:seeds}
\end{figure}
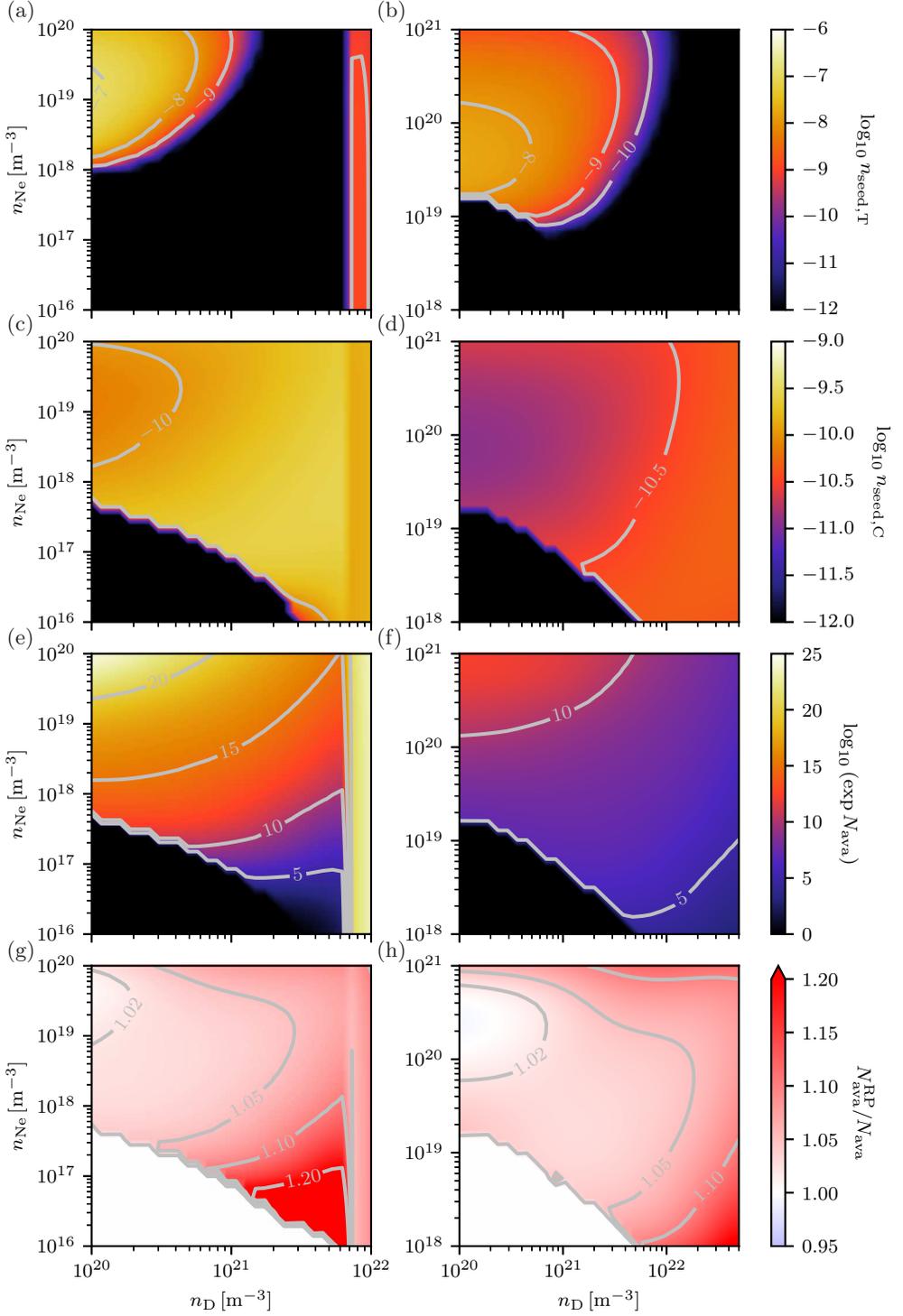

To better understand the contribution from each generation mechanism, inequality \eqref{eq:criterion} can be decomposed into its constituents, namely the tritium decay seed, the Compton scattering seed, and the avalanche gain factor \(\exp{(N_\mathrm{ava})}\), all of which are illustrated in figure \ref{fig:seeds} for both ITER and SPARC. For comparison, the ratio \(N_\mathrm{ava}^\mathrm{RP}/N_\mathrm{ava}\) is plotted as well. The quantitative values differ between the two devices but are qualitatively similar for a given source. Very few runaway electrons are generated in the lower left quadrant, above temperatures of \(\gtrsim \SI{50}{\electronvolt}\). Note that additional physics mechanisms not included in the model presented here, such as those related to a vertical displacement event (including wall impurities entering the plasma), could lead to runaway production even in this parameter region. Furthermore, in addition to minimizing the potential damage caused by runaway electrons, a disruption mitigation system must also manage thermal loads and electromagnetic forces --- constraints that are not considered in the present work.

Moving up in either neon or deuterium concentration, the equilibrium temperature drops, and a large electric field is induced. At higher deuterium concentrations, the free electron density (and thus the critical electric field) increases, yielding a lower value of the normalized electric field \(E\). In the neon rich top left quadrant, the critical electric field remains relatively low, and \(E\) thereby becomes large. This results in a low critical momentum that enables tritium beta electrons to run away. This also coincides with strong avalanching due to the combination of a strong electric field (relative to \(E_\rmc\)) and a large number of bound target electrons, which contribute only weakly to the collisional drag. The result is a substantial generation of runaway electrons due to the combination of a large tritium seed (see figure~\ref{fig:seeds}a--b) and strong avalanching (see figure~\ref{fig:seeds}e--f). Note that the low critical momentum and short thermal quench time can also give rise to a high hot-tail generation rate, which may be greater than the tritium seed generation. But again, hot-tail generation is not considered in this work. 

In this region of parameter space, the Compton scattering seed (figure~\ref{fig:seeds}c--d) is a couple of orders of magnitude weaker than the tritium seed and does not contribute significantly. However, the Compton seed is relatively insensitive to the critical energy and is therefore present everywhere in the considered density space where the temperature is sufficiently low. In fact, going to high deuterium densities enhances the Compton seed due to the larger number of available target electrons.

In ITER, the avalanche gain is sufficiently strong that the Compton seed alone can be amplified into a large runaway current. Therefore, the region in density space where significant runaway occurs is limited by the balance between the Compton seed and the avalanche gain. In SPARC, the avalanche gain is approximately five to ten orders of magnitude weaker than in ITER, and the Compton seed is not sufficiently large to give rise to a significant runaway electron current according to this model. However, this may change if a larger value is chosen for \(\Gamma_\mathrm{flux}\) or if a different value for the current density is chosen. In this work, the current density is taken to be the average current density \(I_\mathrm{p0}/(\upi a^2)\), which is equivalent to a flat current density profile. The current density can also be chosen as the on-axis value of some peaked profile. Using the current density profile below in \S \ref{sec:1D} as an example, the on-axis current density is approximately 40\% higher than the average current density, yielding an avalanche gain that is approximately five orders of magnitude higher in ITER. This would be in line with the maximum avalanche gain reported by, for example, \citet{Wang2024}.

The contribution from the tritium seed is illustrated in figure \ref{fig:criterion}, where, in addition to the full nuclear seed, \(\mathcal{Z}\) is also evaluated using only the tritium seed, indicating where large runaway currents can be expected if the Compton seed is neglected. In the ITER case in figure \ref{fig:criterion}a, this significantly reduces the extent of the region of significant runaway current, indicating that the dominant seed near the solid boundary is indeed the Compton seed. In the SPARC case in figure \ref{fig:criterion}b, however, the dotted contour overlaps with the solid contour, indicating that tritium beta decay is the dominant nuclear seed wherever large runaway currents can be expected. 

In addition, it is noteworthy that for ITER-relevant parameters and deuterium densities \(\gtrsim \SI{7e21}{\per\meter\cubed}\), the plasma temperature drops to such an extent that deuterium recombines, leading to a strong inductive electric field and a large runaway electron current. This is consistent with previous findings in e.g. \citet{Vallhagen2020} and \citet{McDevitt2023}. 

The analytical version and the slightly more detailed semi-analytical version of \(\mathcal{Z}\) make similar predictions regarding where in density space large runaway currents can be expected, although the analytical formulation tends to be a little more conservative. This stems from \(N_\mathrm{ava}^\mathrm{RP}\) being slightly larger than \(N_\mathrm{ava}\) in a plasma consisting of mostly ionized hydrogen and small amounts of weakly ionized neon or argon, as shown in figures \ref{fig:seeds}g and \ref{fig:seeds}h. The reason for the good agreement between the analytical and semi-analytical versions of the criterion is that where the criterion is valid, i.e. where \(n\lesssim 1\), the factor \(\sqrt{4 \nubar{s}(p_\star)^2 + \nubar{s}(p_\star) \nubar{D}(p_\star) }\) is close to its completely screened limit, and the correction for partially ionized impurities is small. In a plasma dominated by weakly ionized high-\(Z\) impurities on the other hand, effects of partial screening become significant and should be treated more carefully. However, injecting high concentrations of impurities is not attractive from a runaway electron mitigation point of view, as the large number of target electrons can give rise to very large runaway electron currents \citep{Hesslow2019}. Hence, this approximation is not a significant limitation of the analytical model.

\subsection{Comparison with 1D fluid simulations}
\label{sec:1D}
As shown above, the criterion \eqref{eq:criterion} is a good surrogate model for a zero-dimensional fluid model. However, there are cases where the radial dynamics are important. To demonstrate this, one-dimensional \textsc{Dream} simulations with radially varying current density and temperature profiles that were allowed to evolve self-consistently were performed. The shapes of the initial current density and temperature profiles are given by
\begin{align}
    \hat{j}(r) & = \left[1 - \left( \frac{r}{a} \right)^2 \right]^{0.41}, \\
    T(r) & = T_0 \left[ 1 - 0.99\left(\frac{r}{a}\right)^2 \right],
\end{align}
where \(T_0 = \SI{20}{\kilo \electronvolt}\) is the on-axis temperature before the thermal quench. The current density profile is normalized such that the current density integrates to the total plasma current \(I_\mathrm{p0}\). At \(t = 0\), a uniform density of neutral neon and deuterium with a temperature of \SI{1}{\electronvolt} is injected. To simulate heat transport losses across the stochastic flux surfaces during the thermal quench, a Rechester-Rosenbluth diffusion coefficient \citep{RechesterRosenbluth1978} is used. The diffusion coefficient is given by \(D\sim v_\mathrm{th}R_0 (\delta B/B)^2\), where \(v_\mathrm{th}\) is the thermal speed, and \(\delta B/B\) is a measure of the magnetic field fluctuations, which in this work was set to \(3\times 10^{-3}\) \citep{Vallhagen2024,Ekmark2025} to obtain a thermal quench time of the order of 1 ms \citep{Hu2021,Sweeney2020}.

Introducing radial profiles and dynamics complicates the comparison between \textsc{Dream} and the criterion \eqref{eq:criterion}, as the local runaway electron production rate is constrained by atomic physics, which, in turn, depends on the local current density. The current density is lower towards the plasma edge, leading to a lower equilibrium temperature and potentially to deuterium recombination, which can result in enhanced avalanche generation. This is exemplified by \citet{Vallhagen2020}, where their ``case 3'' corresponds to a scenario in which the temperature in the outer part of the plasma is initially substantially lower than that in the core, leading to an off-axis runaway current. This type of behaviour is not captured by the reduced zero-dimensional model presented in this work. 

\begin{figure}
  \centering
  \includegraphics{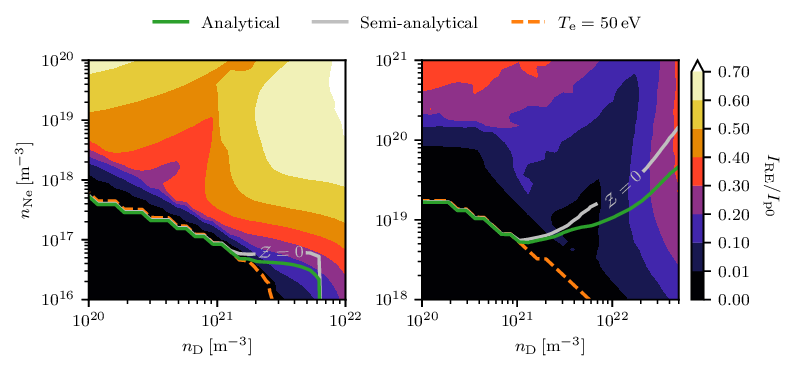}
  \put(-375,153){(a)}
  \put(-215,153){(b)}
  \caption{\textsc{Dream} simulations including radial profiles (filled contours) compared to inequality \eqref{eq:criterion} (green and grey contours for analytical and semi-analytical expressions, respectively) using (a) ITER-relevant and (b) SPARC-relevant parameters. In ITER, the criterion \(\mathcal{Z} = 0\) approximately delineates regions in parameter space where a significant fraction of the Ohmic current is converted into a runaway electron current. In SPARC, the criterion does not capture the off-axis runaway electron generation in the high \(n_\mathrm{D}\) regime. The orange dashed contour indicates the \SI{50}{\electronvolt} isotherm to illustrate how the equilibrium temperature limits the runaway electron generation. Note the nonlinearity in the lower part of the colour map.}
  \label{fig:criterion1D}
\end{figure}

In light of this, the criterion can be adapted to the one-dimensional fluid simulations by using a lower value of \(j_0\) in the power balance \eqref{eq:OhmicRadiativeBalance} to better represent the cold outer plasma. For example, the atomic physics can be evaluated using the average current density \(I_\mathrm{p0}/(\upi a^2)\), whereas the runaway electron dynamics are determined using the on-axis current density. Overall, this makes the criterion more conservative compared to evaluating both using the on-axis current density, since a lower current density leads to a lower equilibrium temperature, which in turn results in a higher electric field and thus a higher runaway electron generation.

In fact, in some cases, the runaway electron generation rate can be viewed as being directly limited by the temperature. This can be seen in figure~\ref{fig:criterion1D}, where the \(T_\rme = \SI{50}{\electronvolt}\) contour is included to illustrate this point. This contour is obtained by solving equations \eqref{eq:OhmicRadiativeBalance}-\eqref{eq:chargeStateDistribution} for \(n_\mathrm{D}\) and \(n_\mathrm{Ne}\) with \(T_\rme = \SI{50}{\electronvolt}\) and \(j_0 = I_\mathrm{p0}/(\upi a^2)\). For deuterium densities below \(\SI{1.5e21}{\per \meter \cubed}\) in ITER and \(\SI{1e21}{\per \meter \cubed}\) in SPARC, the contour \(\mathcal{Z} = 0\) coincides with the contour \(T_\rme = \SI{50}{\electronvolt}\), indicating that for higher temperatures, the critical momentum is sufficiently large to prohibit any significant runaway electron generation. Increasing or decreasing the current density used for the evaluation of the equilibrium temperature therefore shifts the contour to higher and lower densities. Choosing an excessively low Ohmic current could therefore lead the criterion to predict significant runaway electron generation in a region where the plasma would otherwise be too hot to produce a runaway electron beam. The remainder of the contour is determined by the balance of the seed and the avalanche gain, and therefore other parameters such as the total electron density, photon flux, tritium density, and perhaps most importantly, the current density used to evaluate the avalanche gain.

For ITER-relevant parameters, a reasonable agreement between the criterion and one-dimensional \textsc{Dream} fluid simulations is found. The exception is for low neon and high deuterium content, where the off-axis runaway current is not captured (see figure~\ref{fig:criterion1D}a). For SPARC-relevant parameters, the contour \(\mathcal{Z} = 0\) approximately delineates the region of significant on-axis runaway electron generation in the top left quadrant of figure~\ref{fig:criterion1D}b. Again, the criterion does not capture the off-axis runaway current that arises at large deuterium concentrations (\(\gtrsim \SI{1e22}{\per \meter \cubed}\)). This is expected, considering that the criterion has no radial resolution.

Recent modelling efforts have highlighted the importance of runaway electron losses due to the scraping-off of the plasma column during a vertical displacement event \citep{MartinSolis2022,Wang2024,Vallhagen2025,Bandaru2025}. This process has the potential to significantly reduce the final runaway current, especially if the current profile is strongly flattened during the thermal quench, making the plasma prone to produce runaways near the edge. The corresponding non-trivial radial dynamics would be difficult to account for within the current simplified model. However, knowing when large runaway currents are expected in the absence of such losses is nevertheless useful, not only for a quick conservative exploration of the parameter space, but also because scrape-off losses are more effective if the runaway current is not too large even without them; otherwise the current decay stops early, strongly braking the scrape-off process \citep{Vallhagen2025}.

\section{Conclusions}
\label{sec:conclusions}
In future fusion experiments and reactors, the primary runaway electron generation can be dominated by nuclear seeds such as tritium beta decay and Compton scattering. 
In this work, analytical approximations are provided for both tritium and Compton seed densities, as well as for the avalanche gain factor. This results in an analytical heuristic for predicting where in parameter space significant runaway electron generation is likely to occur, based solely on pre-disruption plasma parameters. In addition, a semi-analytical version of the criterion is provided, with a more detailed description of partial screening.



Both the analytical and semi-analytical criteria are found to delineate regions in density space where significant runaway electron generation can be expected for both ITER and SPARC relevant parameters when validated by zero-dimensional fluid simulations conducted with \textsc{Dream}. Compared to one-dimensional fluid simulations, the criterion shows agreement with \textsc{Dream} in large parts of parameter space but does not necessarily capture off-axis runaway beams that are more likely to occur at very high deuterium concentrations. This is particularly evident in SPARC, where the avalanching is weaker.

It is also found that, whenever present, the tritium seed tends to be larger than the Compton seed. Furthermore, the tritium seed correlates with strong avalanching. The Compton seed is less sensitive to the electric field, and the most significant marker for a large Compton seed is the number of target electrons in the plasma. However, note that here the photon flux is set to be constant in time and reduced by a factor of a thousand compared to the values provided in table \ref{tab:Gamma}. 

The close agreement between the analytical and semi-analytical formulations is an indication that, in the parameter ranges considered in this work, a detailed description of the effects of partial screening is not crucial; analytical expressions can be used for simplified analysis at a lower computational cost. Both formulations of the criterion can be evaluated rapidly, making it a useful tool in integrated modelling, system codes, or large parameter scans.

\section*{Acknowledgements}
The authors are grateful to O.~Vallhagen and M.~Hoppe, and E.~Nardon for fruitful discussions.  
\section*{Funding}
The work was supported by the Swedish Research Council (Dnr.~2021-03943 and 2022-02862), and by the Knut and Alice Wallenberg foundation (Dnr. 2022-0087 and 2023-0249). The work has been partly carried out within the framework of the EUROfusion Consortium, funded by the European Union via the Euratom Research and Training Programme (Grant Agreement No 101052200 — EUROfusion). Views and opinions expressed are however those of the authors only and do not necessarily reflect those of the European Union or the European Commission. Neither the European Union nor the European Commission can be held responsible for them. 
\section*{Declaration of interests}
The authors report no conflict of interest.

\bibliographystyle{jpp}

\bibliography{references}

\end{document}